\documentclass[a4paper,11pt]{article}
\pdfoutput=1 
\usepackage{jinstpub} 
\title{\boldmath Strategies to reduce the environmental impact in the MRPC array of the EEE experiment}
\author[1,2,*]{M.P.~Panetta,\note[*]{Corresponding author.}}
\author[3,4]{M.~Abbrescia}
\author[1,5]{C.~Avanzini}
\author[1,5,6]{ L.~Baldini}
\author[1,7]{ R.~Baldini~Ferroli}
\author[1,5,6]{G.~Batignani}
\author[1,8,9]{ M.~Battaglieri}
\author[1,10,11]{ S.~Boi}
\author[12]{E.~Bossini}
\author[1,13,14]{ F.~Carnesecchi}
\author[1,11]{ C.~Cical\`{o}}
\author[1,12,13]{L.~Cifarelli}
\author[1]{ F.~Coccetti}
\author[1,15]{ E.~Coccia}
\author[1,2]{A.~Corvaglia}
\author[16,17]{ D.~De~Gruttola}
\author[16,17]{S.~De~Pasquale}
\author[1,7]{F.~Fabbri}
\author[14]{D.~Falchieri}
\author[1,18,19]{ L.~Galante}
\author[1,14]{M.~Garbini}
\author[9]{ G.~Gemme}
\author[1,20]{ I.~Gnesi}
\author[1,9]{S.~Grazzi}
  \author[1,12,14]{D.~Hatzifotiadou}
  \author[1,21,22]{ P.~La~Rocca}
  \author[23]{ Z.~Liu}
  \author[24]{ L.~Lombardo}
  \author[1,21,25]{ G.~Mandaglio}   
 \author[26]{G.~Maron}
  \author[4]{ M.~N.~Mazziotta}
\author[10,11]{A.~Mulliri}
\author[1,14]{ R.~Nania}
\author[1,14]{ F.~Noferini}
\author[27]{F.~Nozzoli}
\author[1,13]{ F.~Palmonari}
\author[2,28]{ M.~Panareo}
\author[6,29]{ R.~Paoletti}
  \author[24]{ M.~Parvis}
\author[1,26]{ C.~Pellegrino}
\author[1,9]{L.~Perasso}
\author[1,14]{O.~Pinazza}
\author[1,21,22]{ C.~Pinto}
\author[1,7]{S.~Pisano}
\author[1,21,22]{ F.~Riggi}
\author[1]{ G.~Righini}
\author[16,17]{C.~Ripoli}
\author[3]{ M.~Rizzi}
\author[1,13,14]{ G.~Sartorelli}
\author[1,14]{E.~Scapparone}
\author[20,30]{ M.~Schioppa}
\author[29]{ A.~Scribano}
\author[1,14]{M.~Selvi}
\author[1,10,11]{ G.~Serri}
\author[9,31]{ S.~Squarcia}
\author[9,31]{M.~Taiuti}
\author[1,5]{ G.~Terreni}
\author[1,9,25]{ A.~Trifir\`{o}}
\author[1,9,25]{M.~Trimarchi}
\author[26]{C.~Vistoli}
\author[17]{ L.~Votano}
\author[1]{M.~C.~S.~Williams}
\author[1,13,14]{ A.~Zichichi}
\author[1]{ R.~Zuyeuski}
 \author{on behalf of EEE collaboration}
\emailAdd{mariapaola.panetta@le.infn.it}
\affiliation[1]{Museo Storico della Fisica e Centro Studi e Ricerche Enrico Fermi, Roma, Italy}
\affiliation[2]{INFN Sezione di Lecce, Lecce, Italy}
\affiliation[3]{Dipartimento Interateneo di Fisica, Universit\`{a} di Bari, Bari, Italy }
\affiliation[4]{INFN Sezione di Bari, Bari, Italy}
\affiliation[5]{INFN Sezione di Pisa, Pisa, Italy}
\affiliation[6]{Dipartimento di Fisica, Universit\`{a} di Pisa, Pisa, Italy }
\affiliation[7]{INFN Laboratori Nazionali di Frascati, Frascati (RM), Italy}
\affiliation[8]{Thomas Jefferson National Accelerator Facility, Newport News, VA 23606, USA}
\affiliation[9]{INFN Sezione di Genova, Genova, Italy}
\affiliation[10]{Dipartimento di Fisica, Universit\`{a} di Cagliari, Cagliari, Italy}
\affiliation[11]{INFN Sezione di Cagliari, Cagliari, Italy }
\affiliation[12]{CERN, Geneva, Switzerland }
\affiliation[13]{Dipartimento di Fisica, Universit\`{a} di Bologna, Bologna, Italy}
\affiliation[14]{INFN Sezione di Bologna, Bologna, Italy }
\affiliation[15]{Gran Sasso Science Institute, Italy}
\affiliation[16]{Dipartimento di Fisica, Universit\`{a} di Salerno, Salerno, Italy}
\affiliation[17]{INFN Gruppo Collegato di Salerno, Salerno, Italy}
\affiliation[18]{Dipartimento di Scienze Applicate e Tecnologia, Politecnico di Torino, Torino, Italy}
\affiliation[19]{INFN Sezione di Torino, Torino, Italy }
\affiliation[20]{INFN Gruppo Collegato di Cosenza, Laboratori Nazionali di Frascati (RM), Italy }
\affiliation[21]{INFN Sezione di Catania, Catania, Italy }
\affiliation[22]{Dipartimento di Fisica e Astronomia, Universit\`{a} di Catania, Catania, Italy }
\affiliation[23]{ICSC World laboratory, Geneva, Switzerland }
\affiliation[24]{Dipartimento di Elettronica e Telecomunicazioni, Politecnico di Torino, Torino, Italy}
\affiliation[25]{Dipartimento di Scienze Matematiche e Informatiche, Scienze Fisiche e Scienze della Terra, Universit\`{a} di Messina, Messina, Italy}
\affiliation[26]{INFN-CNAF, Bologna, Italy }
\affiliation[27]{INFN Trento Institute for Fundamental Physics and Applications, Trento, Italy}
\affiliation[28]{Dipartimento di Matematica e Fisica, Universit\`{a} del Salento, Lecce, Italy}
\affiliation[29]{Dipartimento di Scienze Fisiche, della Terra e dell'Ambiente, Universit\`{a} di Siena, Siena, Italy}
\affiliation[30]{Dipartimento di Fisica, Universit\`{a} della Calabria, Rende (CS), Italy }
\affiliation[31]{Dipartimento di Fisica, Universit\`{a} di Genova, Genova, Italy}
\abstract{The Extreme Energy Events (EEE) Project employs Multi-gap Resistive Plate Chambers (MRPCs) for studying the secondary cosmic ray muons in Extensive Air Showers. The array consists of~about~60 tracking detectors, sparse on Italian territory and at CERN. The MRPCs are flowed with a gas mixture based on  $C_2H_2F_4$ and $SF_6$, both of which are fluorinated greenhouse gases with a high environmental impact on the atmosphere. Due to the restrictions imposed by the
European Union, these gases are being phased out of production and their cost is largely increasing. 
The EEE Collaboration started a campaign to reduce the gas emission from its array 
with the aim of containing costs and decreasing the experiment global warming impact.
One method is to reduce the gas rate in each EEE detector. Another is to develop a gas recirculation system, whose prototype has been installed at one of the EEE stations located  at CERN.
Jointly a parallel strategy is focused on searching for environmental friendly gas mixtures which are able to substitute the standard mixture without affecting the MRPC performance.
An overview and the first results are presented here.}
\begin{document}
\maketitle
\flushbottom
\section{Introduction}
\label{sec:intro}
The EEE experiment \cite{Zic} is a project by Centro Fermi \cite{CF} (Museo Storico della Fisica e Centro
Studi e Ricerche \textit{Enrico Fermi}), in collaboration with INFN, CERN and several Italian universities.
The project uses the largest array based on MRPC detectors for the study of Extensive Air Showers (EAS) produced by the interaction of high energy primary cosmic rays in the Earth's atmosphere. 
The array extension, characteristics and performance allow to study several cosmic-ray related topics,
from individual EAS and the coincidence between two different correlated air showers, to upward-going events and flux variations of the secondary cosmic rays \cite{search, Up, forbush}. 
It is organized as a network of independent cosmic ray telescopes, spread all over the Italian territory ($\approx 3\times10^5$~km$^2$) and at CERN, 
where the information collected from each station is synchronised by means of GPS technology.

An EEE telescope is made of 3 layers of MRPC at a relative distance of about 50~cm flowed 
with a standard gas mixture of $98\%$ tetrafluoroethane ($C_2H_2F_4$, whose commercial name is R134a) and $2 \% \;$sulfur hexafluoride ($SF_6$) at a continuous flow rate of $2-3\;$l/h ($\approx$1 refill/day).
Due to the large number of telescopes, the gas emission in atmosphere from the array is of the order of $10^6\,$l/year. 
Moreover, $C_2H_2F_4$ and $SF_6$ are characterized by a Global Warming Power (GWP) of 1430 and 23900, respectively. 
The GWP is a measurement of the contribution of a certain gas to the greenhouse effect, normalized to the one from an equal quantity of $CO_2$ (hence $GWP_{CO_2}=\,$1). 
New European regulations, following the Kyoto protocol, prohibit the use of gas or gas mixtures with a $GWP\,>150$ for industrial and commercial uses. 
These greenhouse gases continue to be available for research purposes, and scientific laboratories are excluded from the restriction, but, due to the reduced interest from industry, their cost has largely increased and in the future they will be progressively phased out of production. 
The gas mixture used in the EEE telescope is characterized by a high GWP value, and therefore the collaboration has started a campaign featuring various strategies in order to reduce the total gas emission in the short term and to employ, later on, new eco-friendly gas mixtures.
 \begin{figure}[!htbp]
\centering
\hspace{-1cm}
\begin{minipage}[t]{.5\linewidth}
\centering
\includegraphics[width=1\textwidth]{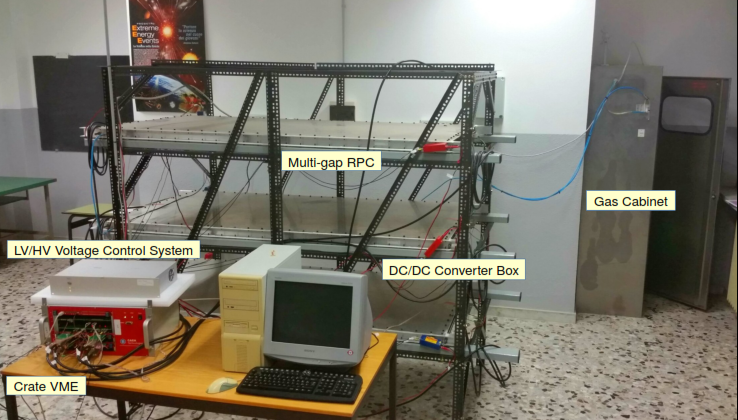} 
\caption{\label{fig:savo}A EEE telescope in Savona, SAVO-02.}
\end{minipage}%
\begin{minipage}[t]{.5\linewidth}
\centering
\includegraphics[width=1\textwidth]{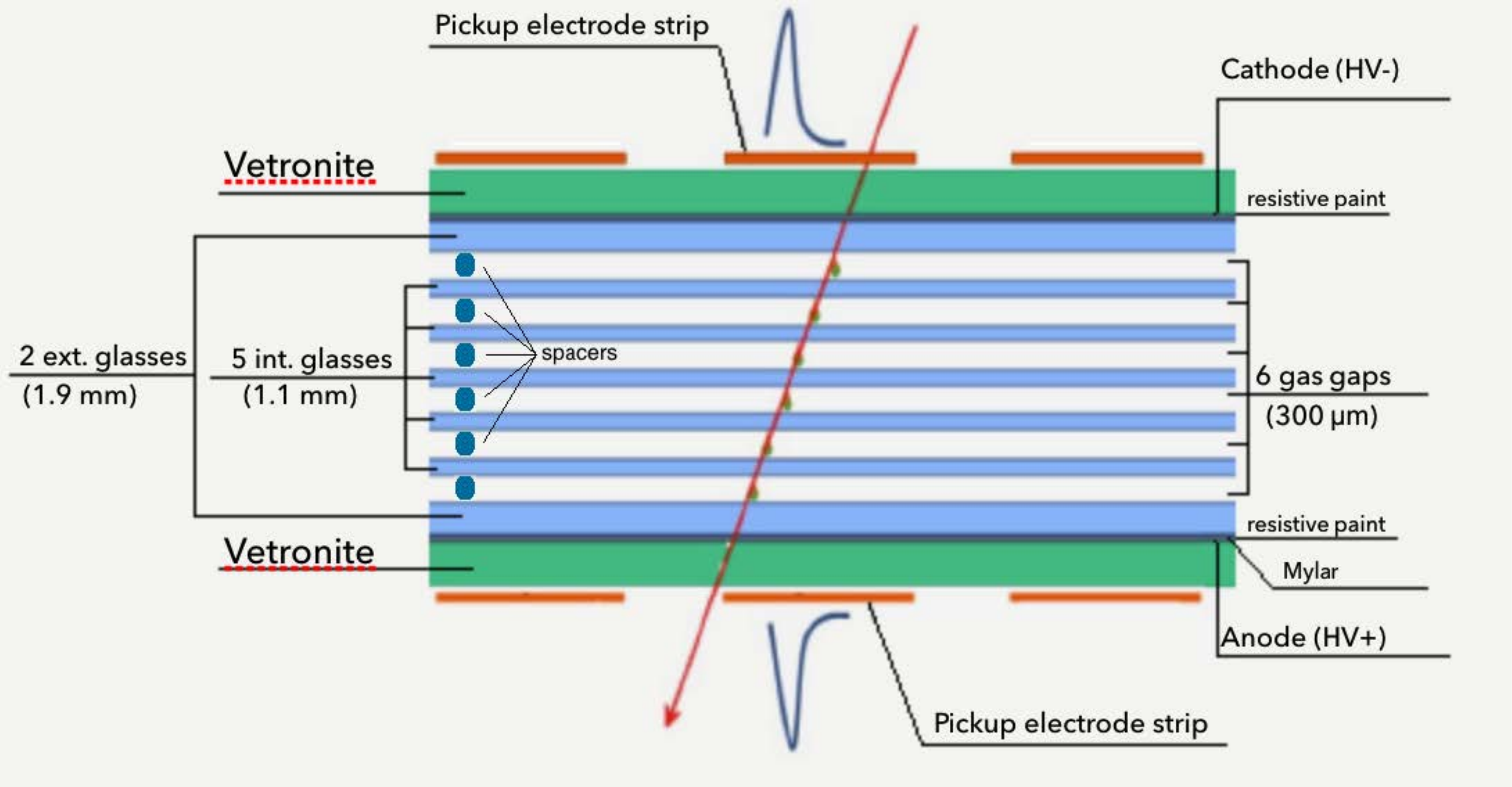} 
\caption{Layout of an EEE MRPC\label{fig:layout}}
\end{minipage}
\end{figure}
\section{The EEE MRPCs}
\label{sec:mrpc}
 A detailed description of the EEE cosmic ray station, shown in Figure~\ref{fig:savo}, and its data acquisition system, has been presented in different papers, as in \cite{costr, Per}, whereas we only introduce few fundamental detector aspects here.
 
 The EEE MRPC is a large detector with an active area of $158\; \times\; 82 \;$cm$^{2}$, shown in Figure~\ref{fig:layout}. The small six gaps, of $300\, \mu$m, are obtained interleaving glass plates by using nylon fishing line.
 The inner glass plates are left electrically floating, and the outer glass plates are coated with resistive paint (whose surface resistivity is about $5M \Omega / \square$), and act as high voltage electrodes.
Two vetronite plates are placed on top and bottom of the outer electrodes, and 24 readout copper strips are glued on both vetronite plates.
Two honeycomb panels reinforce the structure which is placed in an aluminum box of $200\; \times\; 100 \;$cm$^{2}$, sealed by  silicone and screws.
MRPC operates in saturated avalanche mode with a nominal voltage around 18 kV, supplied by two DC/DC converters. 
In the new MRPCs, built after 2017, the gap width has been reduced from $300 \mu$m to $250 \mu$m in order to reach 
a full efficiency at a lower HV value (namely around 16 kV).

EEE telescopes are able to identify the tracks and to measure the time of flight of the cosmic muons crossing through them. 
Muon impact points are indentified in each MRPC, by means of the fired strips, using the difference of the signal arrival times at the strip edges. 
\section{Gas flow reduction}
Each EEE telescope station is equipped with an independent gas line, two bottles and a gas mixer (usually located in a gas cabinet like the one shown in Figure~\ref{fig:savo}), which injects the gas mixture in the 3 MRPCs at a continuous flow rate $\geq 2\,$l/h.
The first step to reduce the impact to the environment from the EEE telescope array is reducing the gas flow in the MRPCs.
However a reduction in the gas flow rate, must be preceded by a campaign aimed at eliminating, as possible, any leak in the gas line, in order to be sure that enough gas mixture really enters the last chamber in the line.

Therefore the EEE collaboration has been carrying out a campaign of gas tightness measurements in the entire MRPC array, 
whose results will be shown here.
\subsection{Tightness test of the MRPCs}
A gas tightness measurement for each MRPC was performed by applying a pressure drop technique, extensively described in \cite{upgrade}.
 \begin{itemize}
    \item[-]A known volume of air ($\approx 1000\,$ml) is gradually injected in the MRPC. The corresponding overpressure $p_c$ is measured (relative to the atmospheric pressure), obtaining a calibration curve $V_c$($p_c$) which can be approximated, under normal condition, by a parabolic function.
    \item[-]After the air injection, the subsequent pressure drop in the chamber $p_d$ is measured during $\approx 1\,$ hour. 
    \item[-] By means of the calibration curve, the volume decrease $V_{d}(t)$ is  evaluated  in the MRPC using the pressure $p_d$, and corrected for each measurement with the relative room values of temperature and pressure. The volume drop curve can be described by the exponential function: $V_{d}(t)~=~V_0~e^{-kt}$.
    \item[-]The MRPC tightness  is obtained as the volume leakage $\frac{dV}{dt}=-k V_d(t')$,  (where $k$ is extracted by the drop curve) at the time $t'$ when the overpressure is $p'_d\,=\,1\,$mbar, $\approx$ the operational overpressure for the EEE telescopes.
   \end{itemize}
A chamber is accepted if the leakage rate at $1\,$mbar is lower than the maximum amount $\frac{dV}{dt}= 0.1 \;$l/h.
 \begin{figure}[!htp]
\centering 
\hspace{-1cm}%
\begin{minipage}[t]{.52\linewidth}
\flushleft 
\includegraphics[width=1. \textwidth]{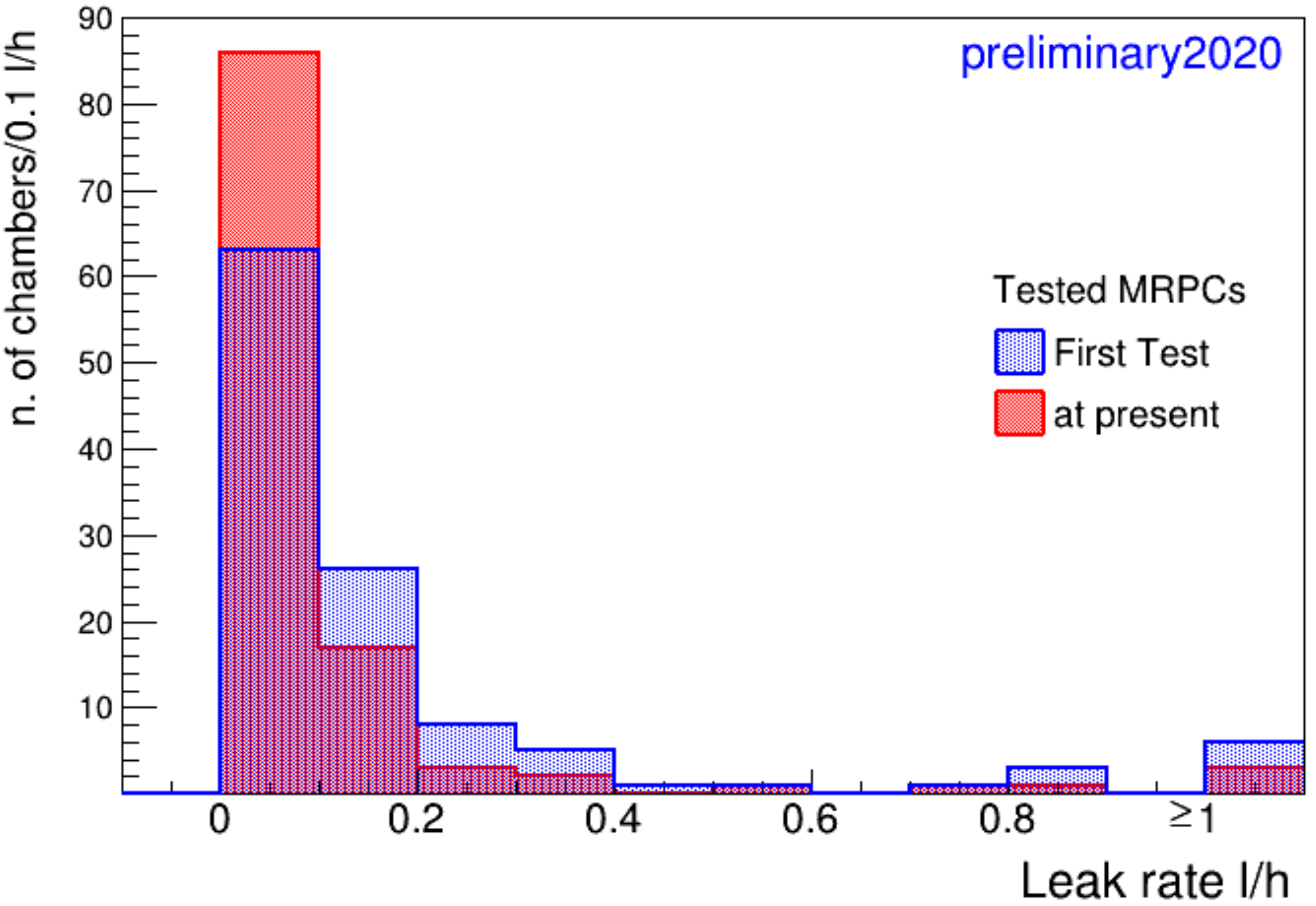} 
\caption{\label{fig:sample} 
 \textit{Blue}: distribution of the gas leak rates for a sample of 120 tested MRPCs. \textit{Red}: the gas leak rates for the same sample, after most of the MRPCs with a leak rate~$>0.1\;$l/h were sealed, repaired and their gas leak measured again. }
\end{minipage}%
\quad
\begin{minipage}[t]{.48\linewidth}
\flushright
\includegraphics[width=1.\textwidth]{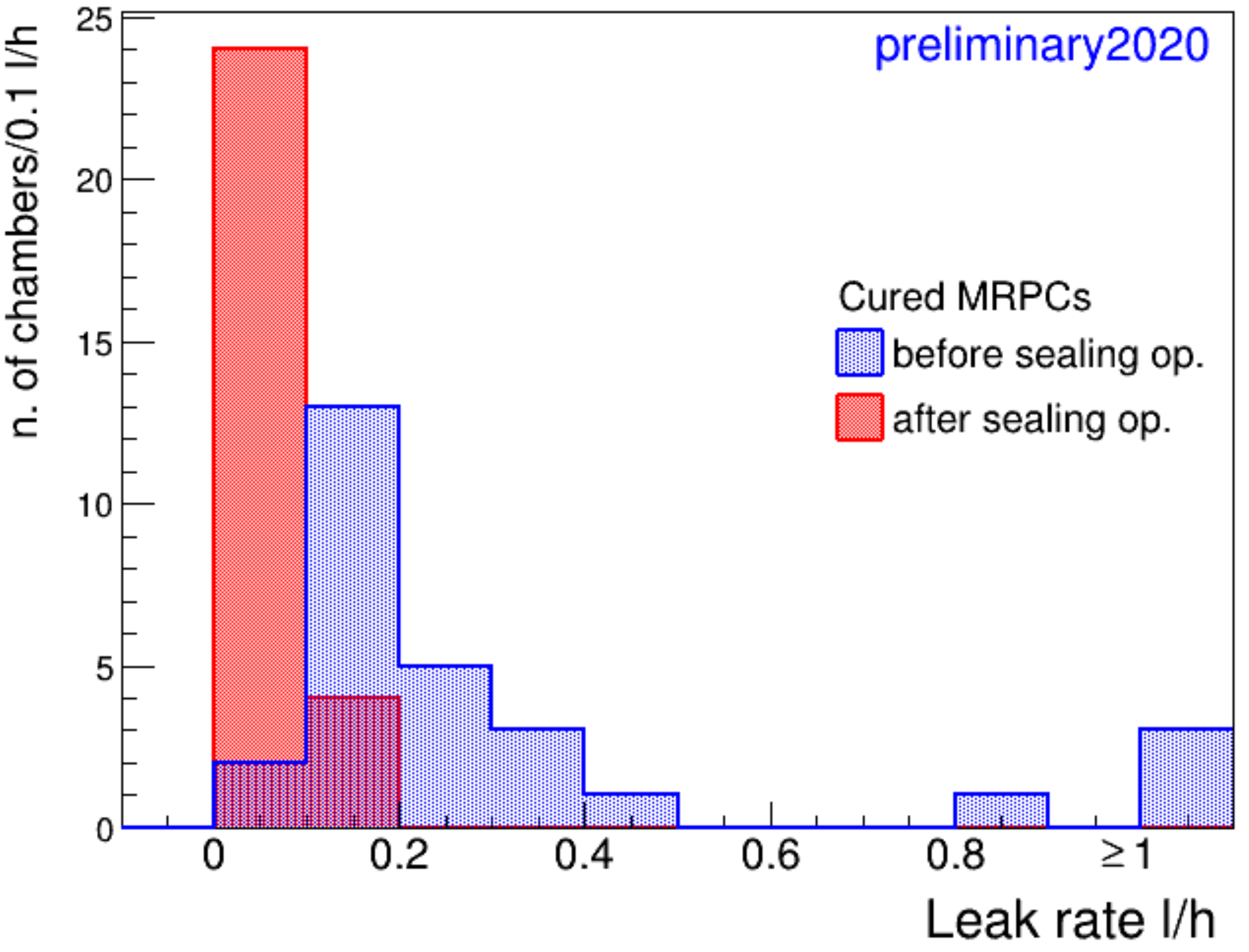} 
\caption{\label{fig:cured} Leak rates distribution for the cured MRPCs (28), before (\textit{blue}) and after (\textit{red}) the sealing operations.} 
\end{minipage}
\end{figure}
 \begin{table}[htbp]
\centering
\caption{\label{tab:ch} MRPCs involved until now in the flow rate reduction campaign, and the campaign results in comparison to the entire array.}
\smallskip
\begin{tabular}{l c c c }
\hline  &  & $n.$ & \textit{ Sample}\\
\hline
MRPCs &         tested  & 120  & 100\% \\
      & leakage $<0.1\,$l/h &  74  & 62\%  \\
      & leakage $>0.1\,$l/h &  46  & 38\%  \\
      &       cured     &  28  & 23\%  \\
\hline      
      \multicolumn{2}{ l }{\textit{At present}   } & $n.$ & \textit{Entire Array}  \\
\hline      
MRPCs      & leakage $<0.1\,$l/h &  112 & 63\% \\
Telescopes & at flow $\approx1\; $l/h &  35 & 60\%\\ 
\hline
\end{tabular}
\end{table}
 The campaign of the gas drop measurements is still ongoing. Until now 40 telescopes (120 MRPCs) were tested, the leak rate results are shown in the distribution in Figure~\ref{fig:sample} (\textit{blue}). 
About 40$\,\%$ of the sample did not pass the tightness test, and it was necessary to find and repair the leaks in the chambers and in their gas line.
HV connectors, gas connectors and gas pipes, screws and MRPC edges were all sealed with silicone.
For the sake of clarity, leak rates before and after the sealing operations are shown in Figure~\ref{fig:cured} for the chambers which failed the first gas tightness test only. 
A summary of the campaign is reported in table \ref{tab:ch}; at present more than 35 telescopes are able to operate at a flow rate of $\approx 1\,$l/h. The performance of these detectors, as time and spatial resolutions and operation stability, has been successfully tested. 
\subsection{Telescopes performance at low flow rate}
 \begin{table}[htbp]
\centering
\caption{\label{tab:res} Time and spatial resolution at different gas flows.}
\smallskip
\begin{tabular}{l c c c c }
\hline 
\textit{Average value}  &  & \textit{flow $\geq 2\;$l/h~~} & \textit{flow $\approx 1 \;$l/h~~} &\textit{ sample (at flow $\approx 1 \, $l/h)}\\
\hline
    Time resolution     & $\sigma_t $ & $ 237 \pm \,  67 \, $ps & $238  \pm \,  40  \, $ps & 22 telescopes\\
Longitudinal resolution & $\sigma_X $ & $1.48 \pm \, 0.04\, $cm & $1.4  \pm \, 0.1 \,  $cm  & 25 telescopes\\
Transversal resolution  & $\sigma_Y $ & $0.92 \pm \, 0.01\, $cm & $0.93 \pm \, 0.05 \,$cm & 25 telescopes\\
\hline      
\end{tabular}
\end{table}
\begin{figure}[!htbp]
 \centering %
 \includegraphics[width=.465\textwidth]{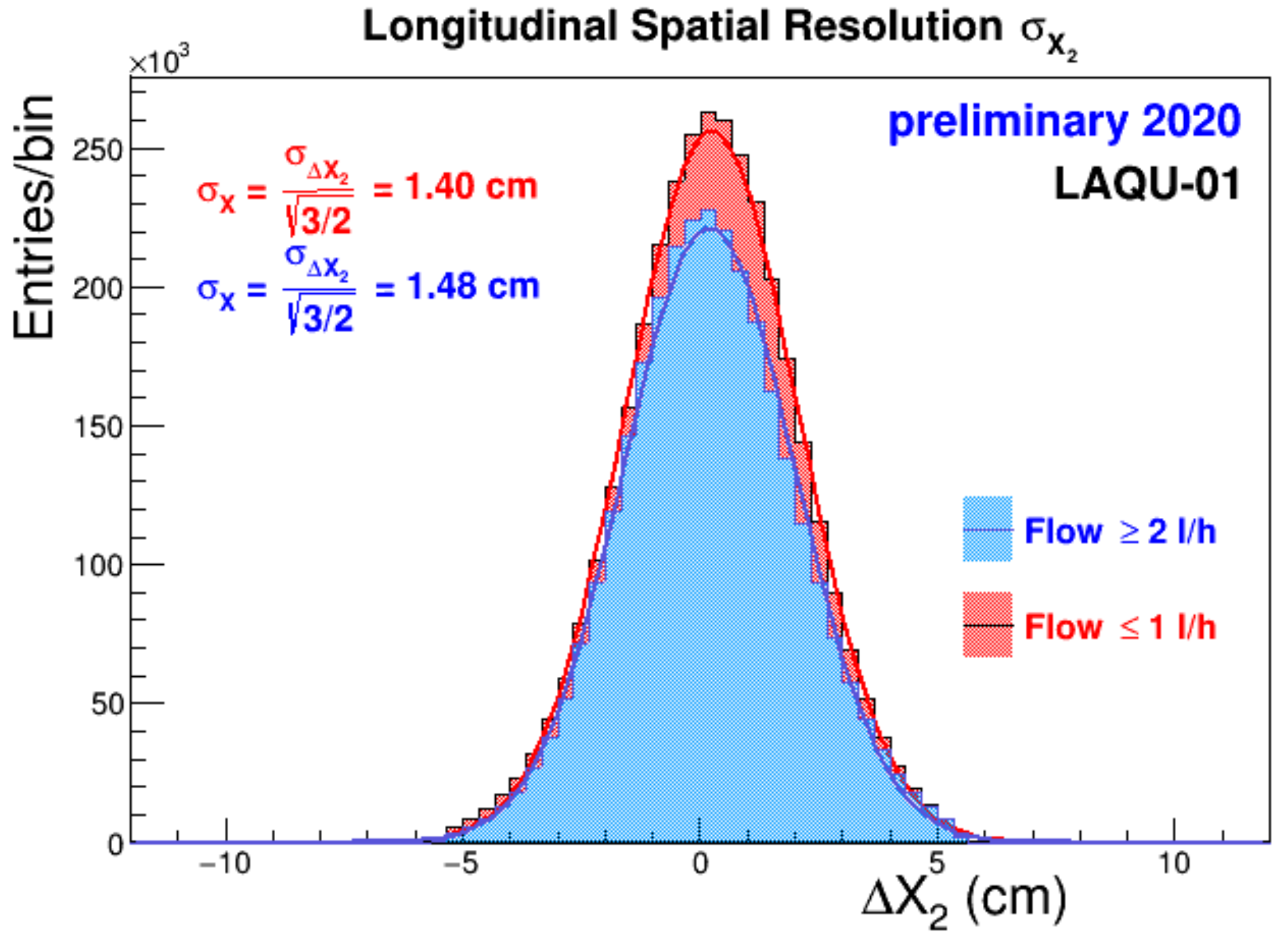}
 \qquad
 \includegraphics[width=.47\textwidth]{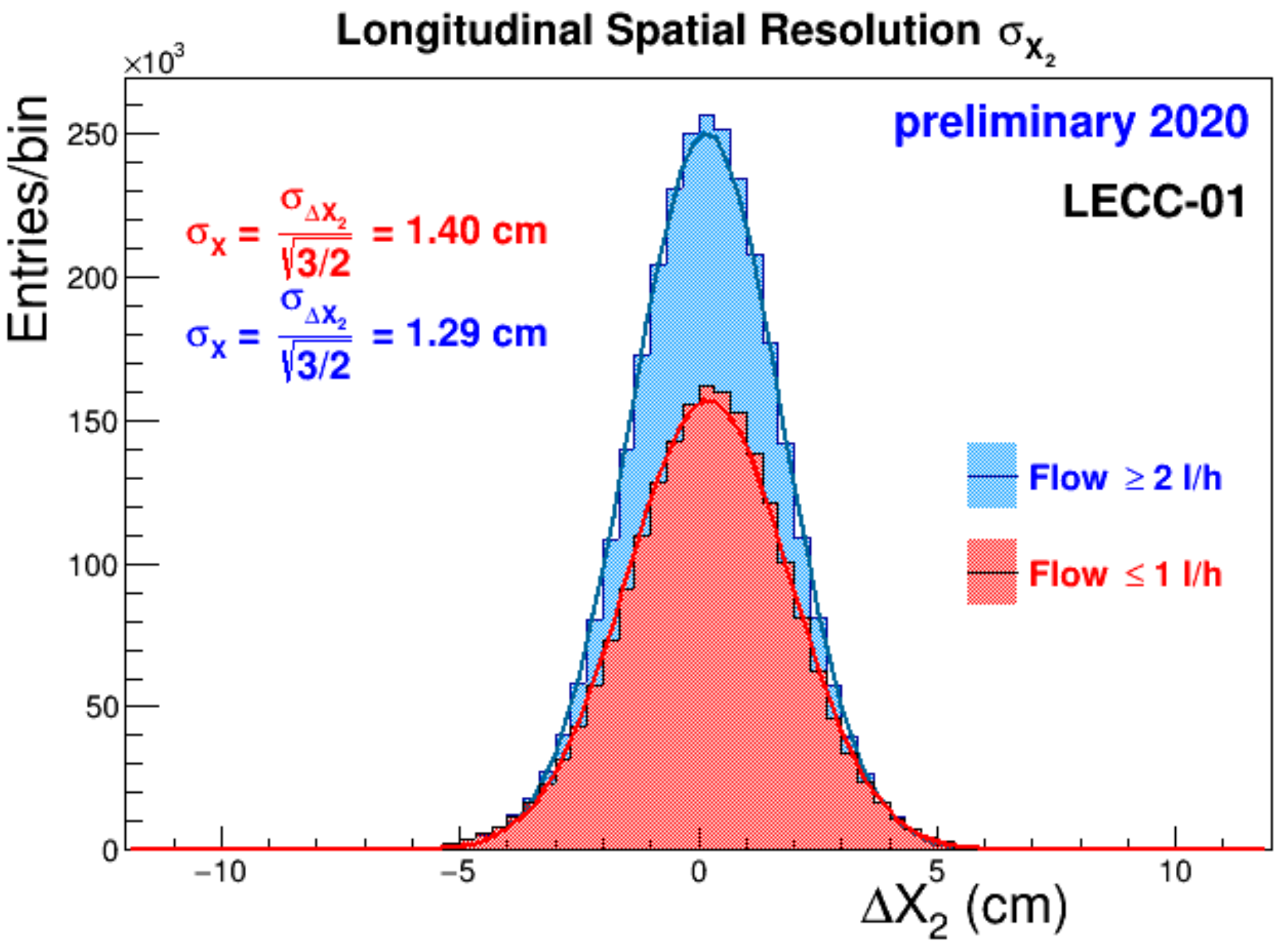}
 \caption{\label{fig:res2} $\Delta X_2$ distributions for two EEE telescopes, LAQU-01 (\textbf{left}) and LECC-01 (\textbf{right}), located in L'Aquila and Lecce respectively. The distribution, in \textit{blue}, is evaluated using particle events with a cut on the tracks for $\chi^2 < 10$, collected in a day during the spring 2019, when the gas flow in the MRPCs was $\geq 2\;$l/h. It is compared with a similar distribution, in \textit{red}, evaluated using data from January 2020 with a gas flow $\approx 1\,$l/h, and after the sealing operations on some MRPCs. The distribution areas are not corrected for the time exposure. The relative longitudinal space resolutions $\sigma_X$ are reported in the canvas.}
 \end{figure} 
A preliminary resolution study for a subsample of 25 detectors has been already described in \cite{Coc}. These results are compared, in table~\ref{tab:res}, with those at low flow rate, obtained before the campaign.  

The spatial and time resolution are measured by extrapolating hit position and hit time in the middle chamber 
from the top and bottom chamber hits and measuring the variance of the residual distribution with respect to extrapolation.
The method is described in more detail in \cite{over,rpc16}. 
As an example the longitudinal resolution with a flow $\geq 2\;$l/h and flow $\approx 1\;$l/h 
for two telescopes, LAQU-01 and LECC-01, is presented in Figure~\ref{fig:res2}.
The resolution is evaluated from the distribution of the variable $\Delta X_2$ that is 
$\Delta X_{middle} = \frac{X_{top}+X_{bot.}}{2}-X_{middle}$. 
 For each distribution the longitudinal spatial resolution $\sigma_X$ is computed as $\sigma_{X}~=~\frac{\sigma_{\Delta X_2}}{\sqrt{3/2}}$ ~\cite{over}, where $\sigma_{\Delta X_2}$ is the distribution variance from the gaussian fit.
The results obtained are compatible, considering the
different operative conditions for the detectors in the two periods of test. 

\begin{figure}[!htbp]
 \begin{flushleft}
 \includegraphics[width=1.0 \textwidth]{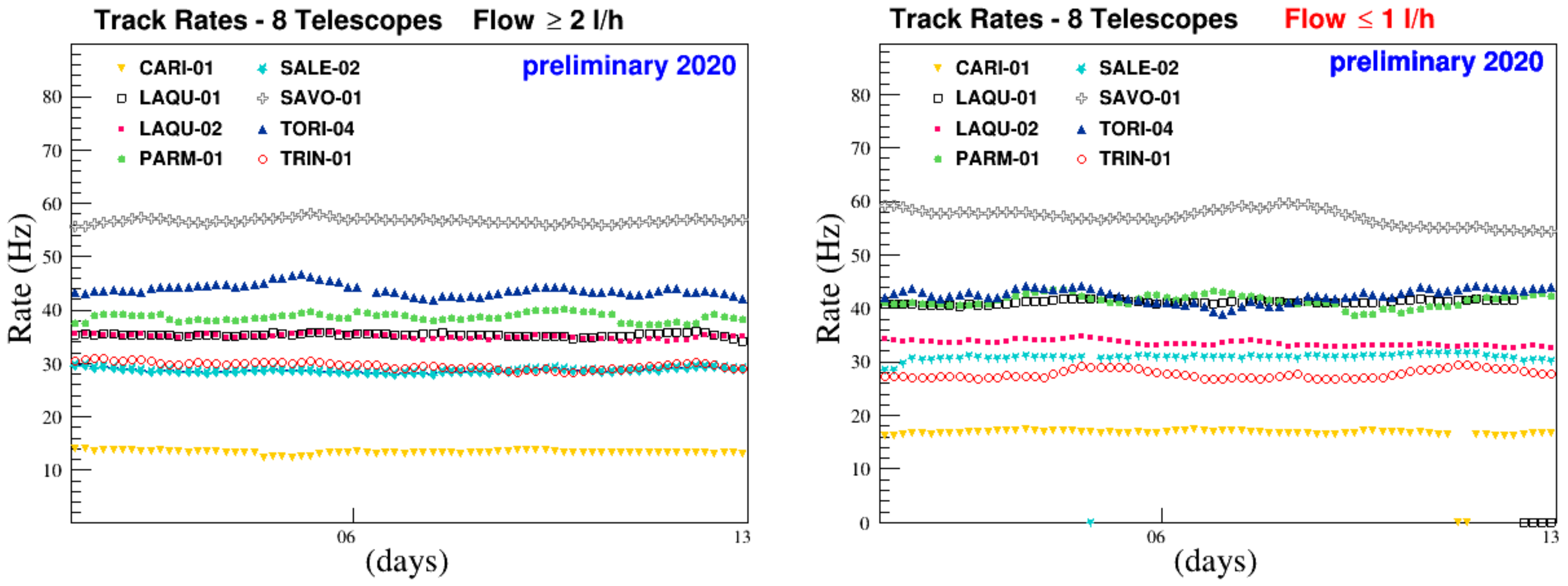}
  \caption{\label{fig:rate} Muon track rates detected in a sample of 8 EEE telescopes, 
  tested before and after the flow reduction campaign. The muon tracks are selected applying a quality cut on their $\chi^2$ value ($\chi^2 < 10$) and a correction for the telescope time exposure is implemented. The rates observed during a test period of 13 days in 2020, with a gas flow through the MRPCs $\approx 1\;$l/h, are presented on the \textbf{right}. A similar data sample taken in spring 2019 when the flow was $\geq 2\;$l/h, is shown on the \textbf{left}. LAQU-01, LAQU-02, PARM-01, SALE-02 and SAVO-01 detectors presented a gas leak $> 0.1\,$l/h in one or more MRPCs, which were repaired.}
   \end{flushleft} 
  \end{figure}
 \begin{figure}[htbp]
  \begin{flushleft}
 \includegraphics[width=1.0\textwidth]{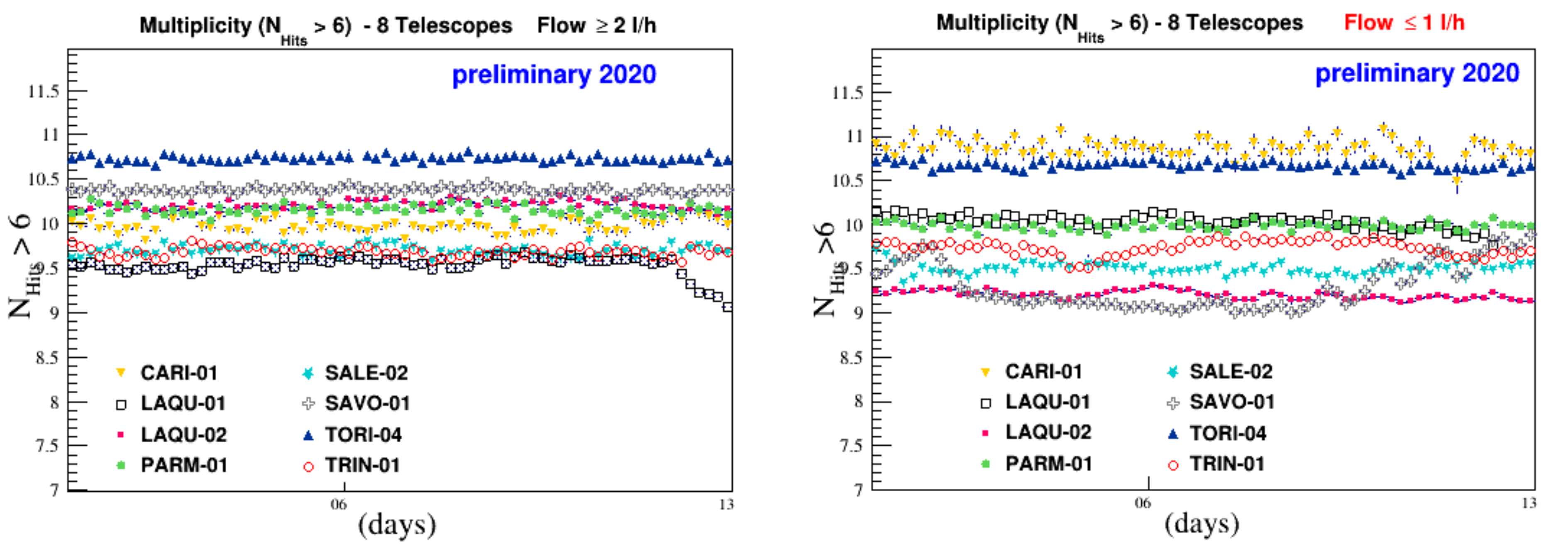} 
 \end{flushleft} 
  \caption{\label{fig:chi2} Multiplicity: total number of hits on the three chambers for each event. The average values, for events with multiplicity $N_{Hits}\; >\;6$ in the telescope, are reported vs. time from 8 telescopes, 
  The trends over the time  are shown for same two data samples of 13 days presented in Figure~\ref{fig:rate}. \textbf{Left}: data collected with gas flow $\geq 2\;$l/h; \textbf{right}: data with flow $\leq 1\;$l/h.}
 \end{figure}
The rate of the particle tracks and the hit multiplicity in the chambers were also monitored for several days to check and control their stability at low flow rate. 
Their values over the time are shown for 8 stations in Figure~\ref{fig:rate} and Figure~\ref{fig:chi2},  compared with an equivalent period at higher flow.
Here the average multiplicity vs. time, namely the average value of the total number of hits on the three chambers, for~$N_{Hits} >6$, is reported. A hit number variation during many hours should be a simple indicator for streamers in the chambers, pointing to a variation in the operating conditions. In fact the multiplicity decrease of LAQU-02 station, where two MRPC showed a high leak rate  during the tightness test ($\approx 1\;$l/h), could be explained by the fact that the leakage was repaired.
The stability of rate and multiplicity, before and after sealing operations
is remarkable, considering the different conditions 
in temperature, external pressure, efficiency fluctuations due to
the time lapse of a year between the two data samples.
\section{Gas Recirculation System}
\begin{figure}[htbp]
 \centering 
 \includegraphics[width=.6\textwidth]{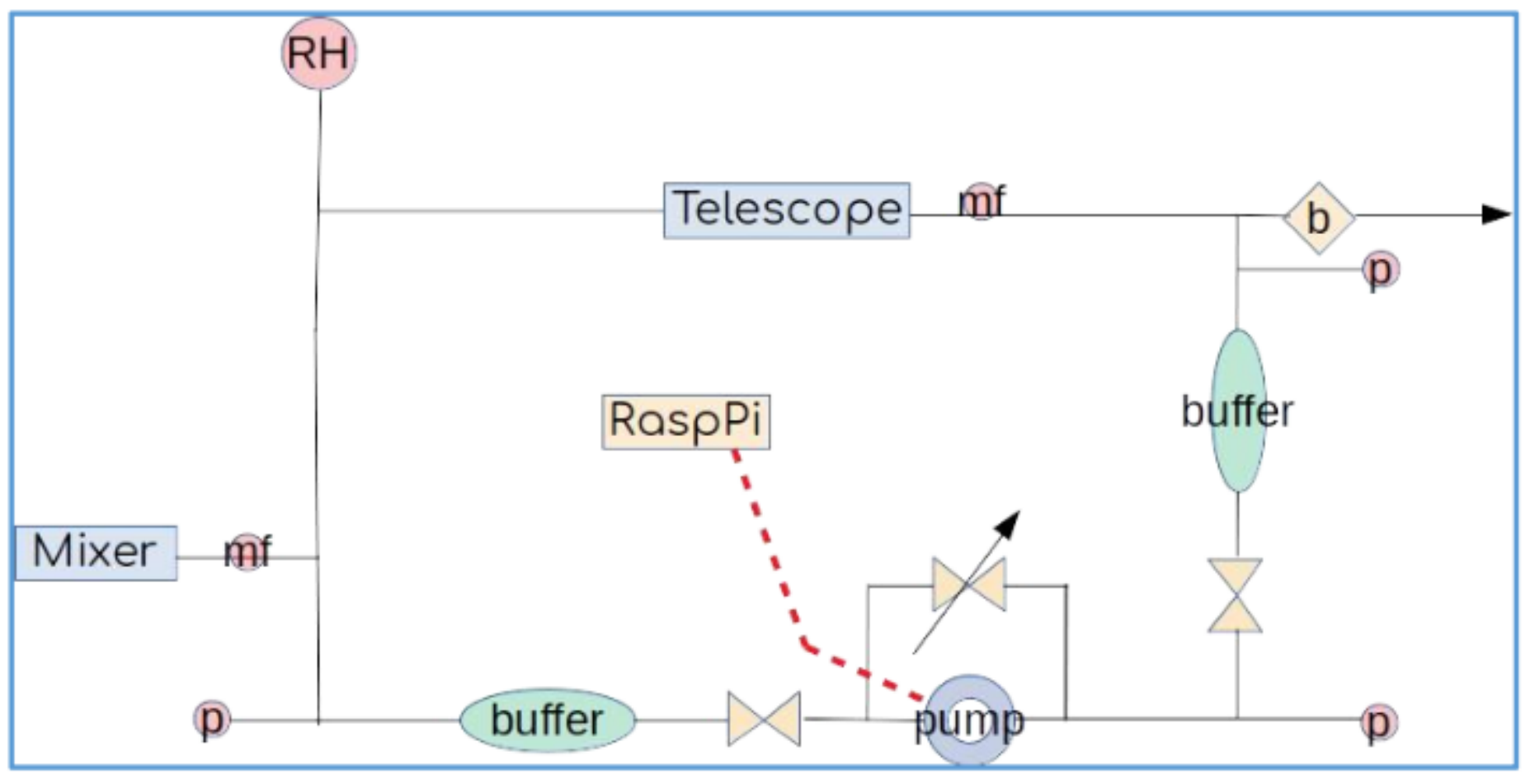}
 \caption{\label{fig:system} Schematic layout for the gas recirculation system. } 
 \end{figure}

The EEE Project needs, for its spread array of independent cosmic ray stations, a simple, compact, cheap, easy-to-use gas recirculation system that should be installed in each station.
A first prototype, in Figure~\ref{fig:system}, developed by the CERN Gas Group, is under test at the CERN-02 telescope.
The gas mixture is collected in the system after being used in the daisy-chain of the 3 MRPCs, and re-injected into the supply lines by means of two buffers and a pump. The humidity is monitored using some dew point probes.
The reused fraction depends on the negative overpressure of the pump branch, monitored by two pressure controls, placed after the mixer (and before the telescope) and after the detectors chain.
Only a small percentage of the gas mixture is sent to the exhaust line and is replaced with fresh gas
mixture coming from the mixer.
At present the prototype is able to re-use a flow fraction of $\approx  60 \, \%$  ($1.8/3 \;$l/h). Some filters are going to be mounted in order to prevent accumulation of impurities, ensuring detectors protection against aging effects. Then the next action  will be to gradually increase the percentage of gas re-used.
%
 %
 %
 \section{Environmentally friendly gas mixtures}
\begin{figure}[htbp]
 \includegraphics[width=.47\textwidth]{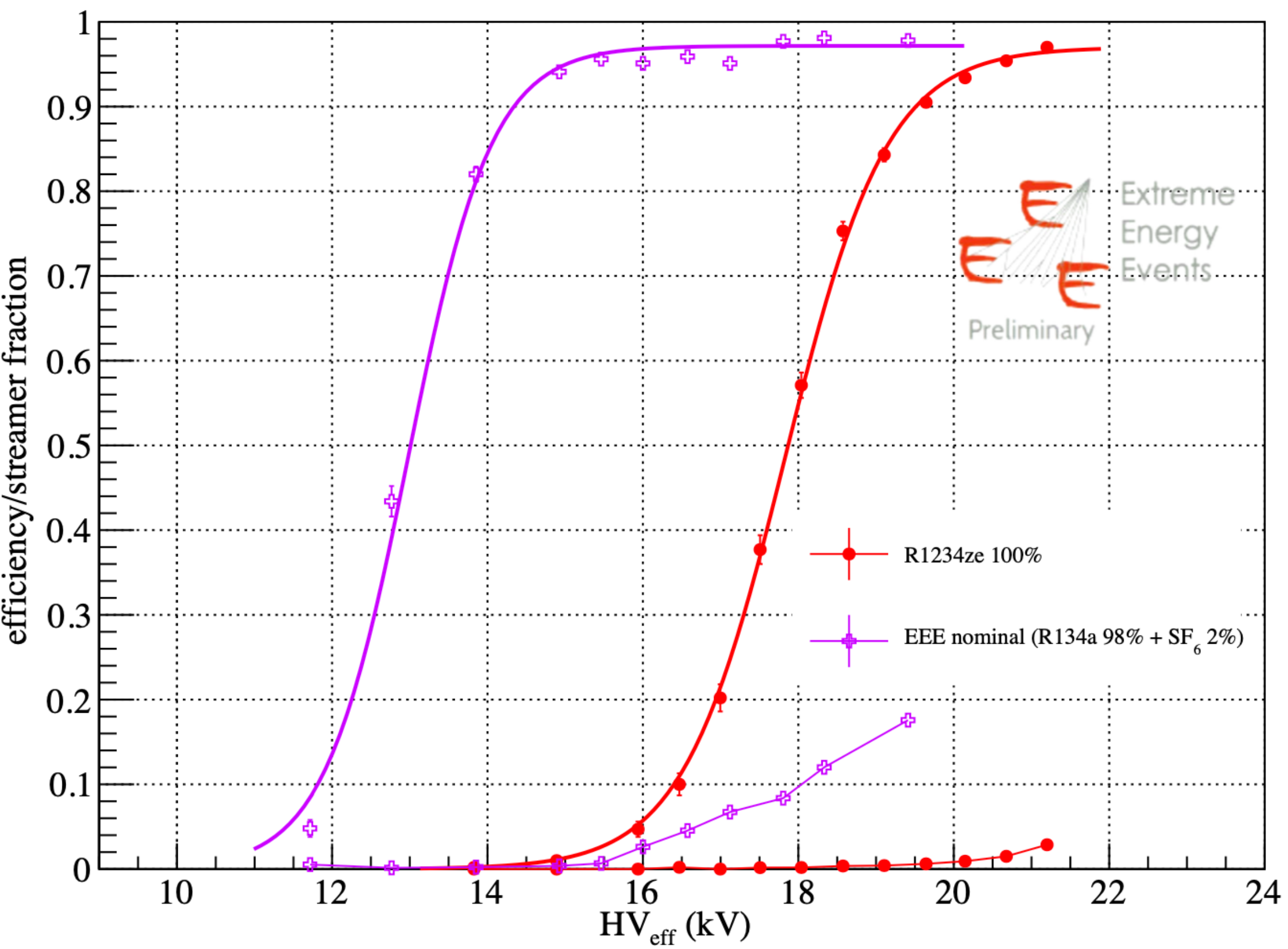}
 \qquad
 \includegraphics[width=.47\textwidth]{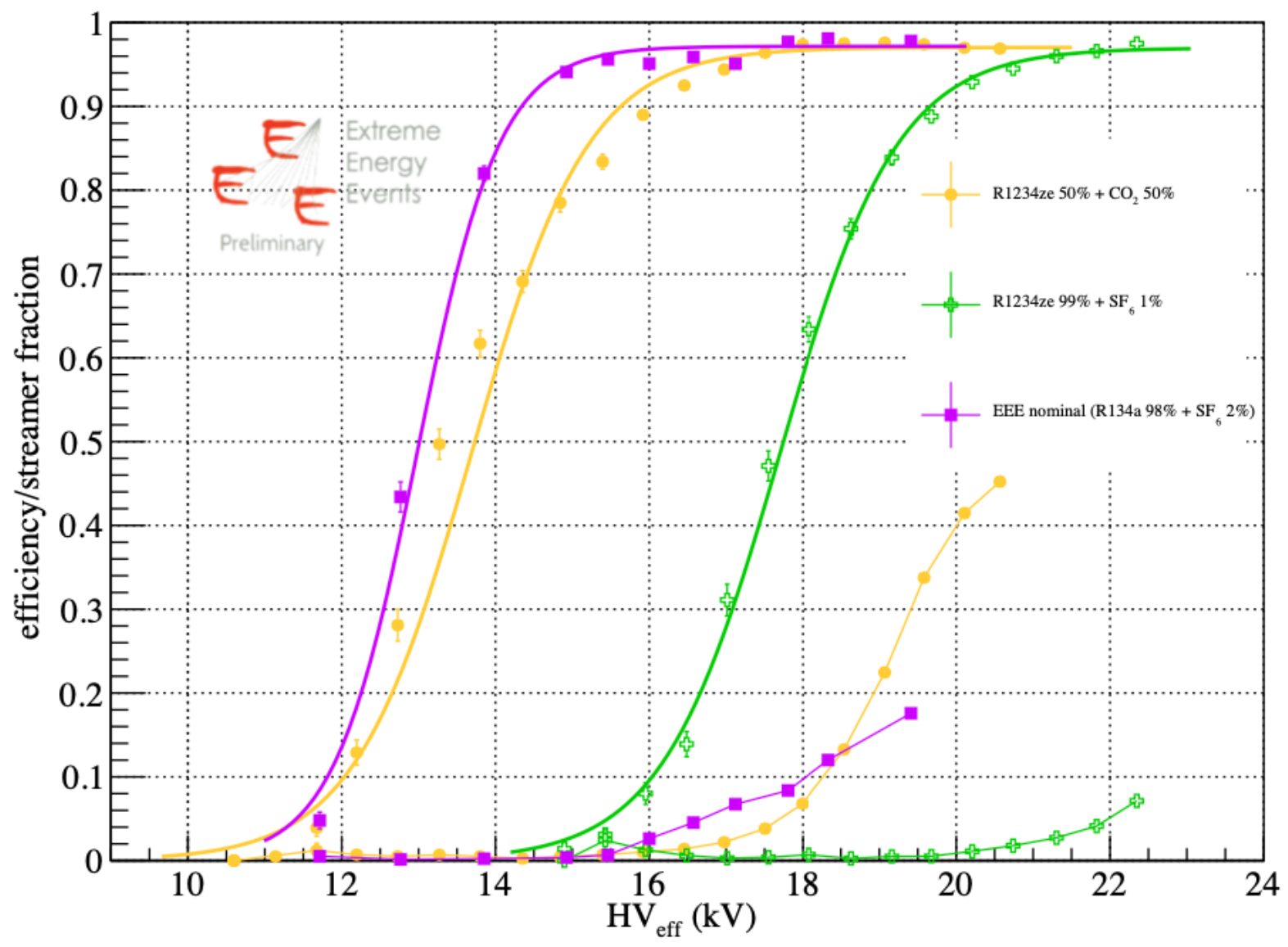}
 \caption{\label{fig:gas2} The MRPC efficiencies and streamer fractions $\bar{s} = \frac{N_{Hits(cluster\geq 5)}}{N_{Tot Hits}}$, as a function of the high voltage, are shown for pure $R1234ze$ \textit{red} on the \textbf{left}; results from $50\% \,R1234ze + 50\% \,CO_2$ (\textit{yellow}) and $99\% \,R1234ze + 1\% \,SF 6$ (\textit{green}) on the \textbf{right}. The nominal mixture values, $98 \% \,R134a +  2\%
 \, SF_6$ (\textit{violet}) are shown for comparison.}
 \end{figure}

Gases in the family of the HydroFluoroOlefin (HFO) have already been identified by the gas detector community as having similar properties to the $C_2H_2F_4$ ($R134$) and a low GWP, and in mixtures with $CO_2$ have already provided
some interesting results when used in RPC detectors \cite{Guida,helium,Baek}.
Thus mixtures based on tetrafluoropropene $C_3H _2 F_4$  ($R1234ze$), an HFO gas, and $CO_2$, using different quenchers, were tested in different percentages also by the EEE community. The procedure, set-up and results have been thoroughly presented in \cite{Eco}. 
The most promising gas mixtures are: 
pure \;$R1234ze~ $ and $~R1234ze \,(50 \%) + CO_2 \,(50 \%)$. \\
The efficiency, current, cluster size etc, as a function of applied high voltage have been studied in a new MRPC (with 250\, $\mu$m  gaps) using cosmic muons. The efficiency plateau is shown in Figure~\ref{fig:gas2}.

Pure $R1234ze$,  with a $GWP =4$ could be a good candidate to substitute
$C_2H_2F_4$, although the gas HV working point, around 21 kV, is
higher than the one with the EEE nominal mixture and close to the upper HV limit supplied by the existing DC/DC converters; 
whereas in the mixture $CO_2 + R1234ze$ the efficiency plateau decreases (because $CO_2$ acts almost like an inert gas), but the streamer percentage increases. In the Figure~\ref{fig:gas2} it is also shown that $SF_6$ acts like a very effective quencher even in small percentage,
but due to its very high GWP value (GWP $\approx$ 23900), its fraction should not exceed 0.5$\%$. 
New mixture tests are ongoing, exploring gases such as $Ar + CO_2$.
\section{Conclusions}
The EEE Project is carrying out a 3-prongs strategy to reduce the greenhouse gases emission already obtaining very positive outcomes.
At present 40 stations are able to take data at low flow, that means the experiment has reached in few months an emission decrease of $34 \%$, without affecting the detectors performance.
 A prototype for the recirculation system is under test and is able to reuse a flow fraction of $\approx 60\%$ with good prospect of improving it.
 Several gas mixtures have been tested in the EEE telescopes, to investigate for replacements of the standard mixture using an environment-friendly one, with encouraging results. 
\acknowledgments
We gratefully acknowledge the CERN Gas Group which developed the gas recirculation system prototype, and has been supporting the EEE collaboration for all the operations of installation and testing of the recirculation unit.
Particular thanks are due to G. Rigoletti and R. Guida. 

\end{document}